\documentclass{emulateapj}
\usepackage[usenames]{color}
\usepackage{graphicx}
\usepackage{amssymb}
\usepackage{amsmath}
\usepackage{times}
\newcommand{\msun}{M_\odot}

\newcommand{\pc}{\mathrm{pc}}
\newcommand{\kpc}{\mathrm{kpc}}
\newcommand{\au}{\mathrm{AU}}
\newcommand{\mbh}{M_\bullet}
\newcommand{\md}{M_\mathrm{d}}
\newcommand{\mc}{M_\mathrm{c}}

\newcommand{\myr}{\mathrm{Myr}}

\newcommand{\sgra}{SgrA{\large$^{\!\star}$}}

\newcommand{\irms}{i_\mathtt{rms}}
\newcommand{\pmin}{p_\mathrm{min}}
\newcommand{\degm}{^\circ}
\newcommand{\kms}{\mathrm{km\,s}^{-1}}
\newcommand{\vesc}{v}
\newcommand{\unit}[1]{#1_\mathrm{u}}

\newcommand{\dist}[1]{n_{#1}}
\newcommand{\Dist}[1]{N_{#1}}

\newcommand{\atan}{\mathrm{arctan}}
\newcommand{\nbody}{$N$-body}
\newcommand{\modela}{{\sf A}}
\newcommand{\modelb}{{\sf B}}
\newcommand{\NEW}[1]{{\bf{}#1}}

\newlength{\itemwidth}
\newcommand{\sublist}[1]{%
 \setlength{\rightmargin}{0cm}%
 \setlength{\itemwidth}{#1}
 \setlength{\leftmargin}{\itemwidth}%
 \setlength{\listparindent}{0em}%
 \setlength{\parsep}{0ex}%
 \setlength{\itemsep}{0.5ex}%
 \setlength{\topsep}{0.5ex}%
 \setlength{\labelsep}{0.5em}%
 \setlength{\labelwidth}{\itemwidth}}
\newcommand{\myitem}[2]{\item[{\makebox[#1][l]{#2}}]}
\shorttitle{The properties of HVSs and S-stars originating from an eccentric disc around a SMBH}
\shortauthors{\v{S}ubr L., Haas J.}
\begin{document}
\title{The properties of hypervelocity stars and S-stars originating from an eccentric
 disc around a supermassive black hole}
\author{Ladislav \v{S}ubr$^{1}$ and Jaroslav Haas$^{1}$}
\thanks{E-mail: subr@sirrah.troja.mff.cuni.cz, haas@sirrah.troja.mff.cuni.cz}
\affil{$^{1}$Charles University in Prague, Faculty of Mathematics and Physics,
Astronomical Institute, V Hole\v{s}ovi\v{c}k\'ach 2, Praha, CZ-18000,
Czech Republic}
\begin{abstract}
Hypervelocity stars (HVSs) that are observed in the Galactic halo, are believed to
be accelerated to large velocities by a process of tidal disruption of binary stars
passing close to a supermassive black hole (SMBH) which resides in the center of the
Galaxy. It is, however, still unclear, where these relatively young stars were born and
which dynamical process pushed them to nearly radial orbits around the SMBH.
In this paper we investigate the possibility that the young binaries originated
from a thin eccentric disc, similar to the one observed in the Galactic center nowadays.
By means of direct {\nbody} simulations, we follow the dynamical evolution of an
initially thin and eccentric disc of stars with a 100\% binary fraction orbiting
around the SMBH. Such a configuration leads to Kozai-Lidov oscillations
of orbital elements, bringing considerable amount of binaries to close vicinity of
the black hole. Subsequent tidal disruption of these binaries accelerates one of
their component to velocities well above the escape velocity from the SMBH
while the second component becomes tightly bound to the SMBH. We describe the main
kinematic properties of the escaping and tightly bound stars within our model
and compare them qualitatively to the properties of the observed HVSs and
S-stars, respectively. The most prominent feature is a
strong anisotropy in the directions of the escaping stars which is observed
for the Galactic HVSs but not explained yet.
\end{abstract}
\newlength{\myfigwidth}
\setlength{\myfigwidth}{0.9\columnwidth}
\section{Introduction}
Hypervelocity stars (HVSs) are observed in the Galactic halo at distances $\gtrsim
50\,\kpc$
from the Galactic center, moving at high velocities $\gtrsim 300\,\kms$, i.e.
they are not gravitationally bound to the Galaxy. To date, about twenty of such
stars were found \citep{Brown2014}, all having spectral type B. 
The particular location and spectral type is, however, just an observational selection.
First, in the Galactic halo, the number density of indigenous stars is low, making the
survey of HVSs efficient. Second, the B-type stars are luminous enough to represent
convenient targets at such large distances. At the same time, these stars are long lived
enough to be able to travel from the Galaxy to the halo. 
It is likely that near future observations will reveal a number of less luminous
HVSs (see \citealt{vickers2015} for one of the first lists of low mass HVS candidates).

S-stars lie, in a certain sense, on the opposite side of the fiducial Galactic scale
from the HVSs -- they are the most tightly bound stars to the supermassive black hole
(SMBH) residing in the center of the Galaxy \citep[e.g.][]{Ghez05,Gillessen09,Meyer12}.
Semi-major axes of their orbits range from $\approx 0.004\,\pc$ to $\approx 0.04\,\pc$.
Similarly to the HVSs, the S-stars are
classified as main sequence B type stars and, similarly to the HVSs, there are
observational limitations which do not allow to detect less luminous main sequence
stars in that region.

It was suggested by \cite{Hills88} already before finding any HVSs or S-stars
that both groups (considering their kinematic state, not necessarily the spectral
type) should exist as a consequence of tidal break-ups of binary stars in the
vicinity of the SMBH. Subsequent works of other authors, particularly after the
discovery of the S-stars and the HVSs, tried to answer the question where these
binary stars were born and which process drove them to orbits plunging below
the tidal break-up radius. Some works considered the binaries to originate
at distances well above $1\,\pc$ from the SMBH and to be transported inwards,
e.g., within a massive young star cluster \citep{Gould03} or scattered due to
relaxation enhanced by massive perturbers \citep{Perets07}.

Another possible source of binaries appears to be quite straightforwardly motivated by
presence of another group of young stars in the Galactic center, which are
observed at radial distances $0.04\,\pc \lesssim r \lesssim 0.4\,\pc$ from the SMBH
\citep[e.g.][]{paumard2006,bartko2009,bartko2010,do2013}. A subset of these stars
apparently orbit around the SMBH in a thin disc \citep[the so-called clockwise
disc, CWS; e.g.][]{Levin03}. Their
binary fraction is currently not well constrained \citep[e.g.][]{Pfuhl2014} either from
the observational data, or from models of star formation in the vicinity of the SMBH.
In spite of the fact that stars from the young stellar disc have moderate values
of orbital eccentricities, their pericenters are several orders of magnitude
above the tidal break-up radius of binaries that can survive in such a dense environment.
Hence, some mechanism has
to come in act to excite the orbital eccentricities in order to transport the binaries
sufficiently close to the SMBH.

\cite{Lockmann08} suggested that the sought mechanism may be the Kozai-Lidov
resonance, induced by external, flattened and axially symmetric source of gravity.
In their particular setup it was a second, highly inclined
stellar disc which was assumed to be formed at the same time in the Galactic center.
Later on, \cite{Madigan09} who studied dynamics of an {\em eccentric\/} disc
of stars around a SMBH noticed that substantial fraction of stellar orbits undergo
extreme oscillations of their eccentricities due to the disc self-gravity.
\cite{Madigan09} interpreted these oscillations as a manifestation of the so-called
eccentric instability and claimed that their occurrence relies on presence of
massive spherical cluster centered on the SMBH. In \cite{Haas15} we have shown,
however, that these oscillations may be interpreted as an imprint of
Kozai-Lidov dynamics in the perturbing gravitational field of the stellar disc itself
which does not require presence of the spherical cluster. \footnote{For further
discussion of the eccentric instability see
J.\ Haas \& L.\ \v{S}ubr (2016), in preparation.}

Both \cite{Lockmann08}
and \cite{Madigan09} suggested that the oscillations of orbital eccentricities
which they observed in their models could contribute to formation of HVSs
and the S-stars. Note, however, that in both works, only orbits of single stars
were integrated in the full {\nbody} setup. Conclusions about the HVSs
and the S-stars formed through the Hills mechanism were based on numbers of single
stars reaching such a vicinity of the SMBH, where a binary would tidally break up.
Another class of numerical models of the HVSs and the S-stars is based on integrations
of individual binaries injected towards the SMBH with arbitrarily given impact
parameters \citep[e.g.][]{Kenyon08,Antonini2010,Zhang13}. Such experiments allow to evaluate statistical
properties of the captured and ejected stars, however, the weak point of the models
lies in poorly constrained input parameters of the orbits around the SMBH prior the
tidal break-up.

In this paper, we make another step forward by including stellar binaries
into {\nbody} models of stellar discs around a SMBH. For the first time,
this setup allows us to make predictions about properties of the HVSs and the S-stars
originating for a young stellar disc in a self-consistent way. Similarly to
\cite{Madigan09} and \cite{Haas15} we consider the parent disc of the young stars
to be formed by aligned eccentric orbits, i.e. the disc itself is capable to
push some of its members to extremely oscillating orbits.
\subsection{The Kozai-Lidov mechanism}
The key process which we assume to drive stellar orbits to highly eccentric state
is the Kozai-Lidov resonance.
This resonance belongs to a class of nearly Keplerian motions around a dominating
central mass which is perturbed by some source of gravity.
In the case of the {\em classical\/} Kozai-Lidov mechanism \citep{Kozai62,Lidov62}
the perturbing gravitational potential is assumed to be axially symmetric, which
implies a conservation of the projection of the angular momentum of the orbiting
body onto the symmetry axis (also called the Kozai-Lidov integral). As the remaining
components of the angular momentum vector may
change, it is possible for the orbit to undergo oscillations from highly eccentric
with a low inclination with respect to the plane of the perturbation to low eccentric
and highly inclined. The classical Kozai-Lidov oscillations were considered by
\cite{Lockmann08} who assumed two mutually highly inclined discs to influence
each other.

When the axial symmetry of the perturbing potential is lost, none of the components of
the angular momentum is an integral of motion any longer. A specific class of
sources of such a perturbing potential relevant for celestial mechanics -- body on
an elliptical orbit around the central mass -- was studied in the literature
\citep[e.g.][]{Ford2000,Katz11,Lithwick11,Li14}. If the deviation from the axial
symmetry is not strong, the component of the angular momentum vector perpendicular
to the plane of the orbit of the perturbing body slowly changes, which leads
to modulations of the classical Kozai-Lidov oscillations. The eccentricity of the
perturbing body motivates us, as well as the previous authors, to call this effect
the {\em eccentric\/} Kozai-Lidov mechanism. In our specific setup with similar
orientations of stellar orbits, the most important variant is then the {\em coplanar}
eccentric Kozai-Lidov mechanism \citep{Li14}.
\section{Model and method}
\label{sec:model}
\subsection{Model of the Galactic center}
\label{sec:model_details}
In order to study the role of the Kozai-Lidov mechanism for production of the HVSs
and the S-stars, we introduce a model of mutually gravitationally
interacting point masses (particles). As there is no explicit characteristic
physical scale in the equations of motion of point masses interacting according
to the Newton's law of gravity, the model may be arbitrarily rescaled provided
the relation $\unit{t}^2 = \unit{r}^3 / G\,\unit{M}$ is fulfilled, where $\unit{t},\;
\unit{r}$ and $\unit{M}$ represent the time, length and mass unit, respectively,
and $G$ stands for the gravitational constant. While in the numerical realization
it is convenient to set $\unit{t} = \unit{r} = G\,\unit{M} = 1$, below we describe
our model scaled to physical units to make it human readable. In particular, we
consider values motivated by the Galactic center: $\unit{M} = 4\times
10^6\,\msun$ and $\unit{r} = 0.004\,\pc$, which implies $\unit{t} = 1.89\,\mathrm{yr}$.
We integrated numerically two different initial settings, labeled {\modela} and
{\modelb}:

\begin{list}{}{\sublist{3em}}
\myitem{2.5em}{(i)} SMBH particle of mass $\mbh = 1\,\unit{M} = 4\times 10^6 \,\msun$
 is initially at rest at the origin of the coordinate system
\myitem{2.5em}{(ii, {\modela})} 2000 stars are drawn randomly from a power-law distribution function
 $\dist{M_\star} \propto M_\star^{-1.5}$ with a lower and an upper boundary
 $1\,\msun$ and $150\,\msun$, respectively, which implies the total mass of the
 disc $\md \approx 24500\,\msun$. This distribution function is
 motivated by a recent analysis of the properties of the young stars in the Galactic
 center done by \cite{Lu13}.
\myitem{2.5em}{(ii, {\modelb})} 2000 stars have equal mass of $M_\star = 12.25\,\msun$, i.e.
 $\md = 24500\,\msun$ as in model {\modela}
\myitem{2.5em}{(iii, {\modela})} All stars are paired to binaries with a preference for the mass ratio
 close to unity. This is motivated by the fact, that massive stars in the Galactic
 field tend to have massive companions. Although this is not true for low-mass stars,
 we used the close to equal-mass pairing for the sake of simplicity. Binary
 semi-major axes were drawn from the \"Opik distribution function \citep{Kobulnicky07},
 $\dist{a_\mathrm{b}} \propto a_\mathrm{b}^{-1}$ with $a_\mathrm{b} \in \langle
 0.1\,\au, 100\,\au \rangle$;
 their initial eccentricity was set to zero and their orbital angular momentum
 parallel to the angular momentum of the disc.
\myitem{2.5em}{(iii, {\modelb})} All stars are paired to binaries with initial semi-major axis
 $a_\mathrm{b} = 10\,\au$, zero eccentricity and the orbital angular momentum
 parallel to the angular momentum of the disc.
\myitem{2.5em}{(iv)} Binaries are placed on elliptic orbits around the SMBH with semi-major
 axes following a power-law distribution function $\dist{a} \propto a^{-1},\;
 a_\mathrm{in} \equiv 0.04\,\pc \leq a \leq a_\mathrm{out} \equiv 0.4\,\pc$.
 Inclinations with respect to some reference plane were drawn from a distribution
 function $\dist{i} \propto \sin i,\; i \in \langle 0, 2.5\degm \rangle$.
 Motivated by the work of \cite{Mapelli12}, who modeled star formation in the
 vicinity of the SMBH from infalling gas cloud, we introduced a non-zero radial
 gradient of eccentricities, namely, we initialized them according to formula
 $e = 0.9 (a - a_\mathrm{in}) / (a_\mathrm{out} - a_\mathrm{in})$. Furthermore,
 in tune with their results, we constructed the orbits around the SMBH such
 that their eccentricity vectors point to a common direction, i.e. the ellipses were
 mutually aligned, not randomly oriented.
\end{list}
If not specified explicitly otherwise, the results presented below correspond
to model {\modela}.

Let us note here that the models presented above do not involve another important
component of the Galactic center -- the spherical cluster of old stars. This is due
to numerical complications (stability and CPU time consumption) which would be
introduced by adding this cluster into the full {\nbody} setup. By means of additional
simple models without binaries, however, we discuss the impact of the
spherical cluster on the results from the main models in Section~\ref{sec:discussion}.
\subsection{Numerical integrator}
We used NBODY6 code \citep{Aarseth03} for the numerical integration of the equations
of motion. We added to the original code routines for logging of beginnings and
endings of regularizations into a binary file. We further introduced an identification
index for the SMBH particle and stored the minimum distance of all other particles
to the one representing the SMBH reached during the integration. Finally, we
altered the decision making algorithm for adding particles
to neighbor lists. In particular,
we weight the standard distance criterion by the mass of the given particle
so that the more massive particles are added to the list even when they are at larger
distances than the lighter ones. The most prominent target was the SMBH particle
which was, due to this modification, a member of the neighbor lists of all other
(star) particles. The modification of the neighbor list influences the integration
and increases its stability.
Among many runtime options of the NBODY6 code, let us specifically mention that
we switched off the internal evolution of the stars as well as the post-Newtonian
corrections to the stellar dynamics.
\subsection{Escapers and hypervelocity stars}
Analysis of the escaping stars in our models was done as follows. First, we
identified all stars that reached a distance of $8\,\pc$ from the SMBH with a velocity
exceeding the escape velocity from the SMBH at that radius ($\approx 66\,\kms$).
For each of these stars we search the list of the regularization events for the last
one including its ID. This event provided us with
information about the ejection time, $t_\mathrm{ej}$, and radius, $r_\mathrm{ej}$.
Finally, we scan the event list file for all events in time interval $\langle
t_\mathrm{ej} - \Delta t, t_\mathrm{ej} \rangle$ involving the respective star.
We count the number of unique stellar IDs that occur in these events in order to
determine the type of ejection mechanism. The time interval $\Delta t$ is taken
to be the greater of the orbital period around the SMBH with semi-major axis equal to
$r_\mathrm{ej}$ and $1 \unit{t}$. We explicitly distinguish the ejection process
involving just two stars -- a tidal disruption of binary -- which we call the
{\em Hills mechanism\/} throughout the rest of the paper. The other ejection
cases that involve more than two stars are called {\em multi-body ejections}.
The method of determination of the ejection
event as well as the number of stars involved depend on the details of the regularization
techniques implemented in the NBODY6 code which are not designed directly
for this purpose. Therefore, there is a small fraction (of the order of one per cent)
of the escaping stars whose $r_\mathrm{ej},\; t_\mathrm{ej}$ and mechanism of
the ejection may be misdetected.

Not all of the stars which reach the distance of $8\,\pc$ with velocity larger
than the escape velocity from the SMBH would also reach the Galactic halo.
In order to present results suitable for a comparison with the observed HVSs,
we performed a reduction of velocities which accounts for the loss of the kinetic
energy in the potential of the Galaxy. In particular, we considered a relatively
simple compound spherical potential of the Galaxy and the SMBH \citep[e.g.][]{Kenyon08},
\begin{eqnarray}
 \Phi(r) &=& \Phi_\mathrm{G}(r) + \Phi_\bullet(r)
 \label{eq:galpot} \\
 &=& 2\pi GCr_\mathrm{c}^2 \left[\frac{2r_\mathrm{c}}{r} \atan{\frac{r}{r_\mathrm{c}}}
 + \ln\left(1 + \frac{r^2}{r_\mathrm{c}^2} \right) \right]
 - \frac{G\mbh}{r}\;,
 \nonumber
\end{eqnarray}
where $C$ and $r_\mathrm{c}$ are
free parameters of the model for which we adopt values $C=1.4\times10^4 \msun
\pc^{-3}$ and $r_\mathrm{c} = 8\,\pc$. Specific kinetic energy of
stars then decreases by a value of $\Delta E_\mathrm{K} \approx 4\times 10^5 (\kms)^2$
when traveling
from $8\,\pc$ to $50\,\kpc$, i.e. from the radius where the Galactic potential starts
to dominate to the typical Galactocentric distance of the observed HVSs.
In the following, we thus distinguish all the {\em stars escaping
from the SMBH\/} at the distance of $8\,\pc$ (or shortly {\em escapers}) and the
{\em hypervelocity stars (HVSs)\/} which
are the subset of escapers capable to reach the galactocentric distance of $50\,\kpc$
(i.e. having velocities greater than $894\,\kms$ at $8\,\pc$).
\subsection{S-stars}
\label{sec:sstars-def}
Formally, we define S-stars in our model to be the former companions of the escaping
stars originating from the Hills mechanism which stay bound to the SMBH. We will
show below that an overwhelming
majority of these stars are located below the inner edge of the parent stellar disc,
i.e. at the places where the observed S-stars are found in the Galactic center. Orbital
elements of the S-stars are evaluated in the rest frame of the SMBH particle.
\section{Results}
\label{sec:results}
\subsection{Properties of the hypervelocity stars}
\label{sec:results_hvs}
We have integrated 150 realizations of model {\modela} up to $t = 4 \times 10^6\unit{t}$
which corresponds to $t\approx 7.6\,\myr$ for scaling $\unit{r} = 0.004\,\pc$ and
$\unit{M} = 4 \times 10^6\,\msun$. On average, we found $34$ escaping stars per
realization out of which approximately $5/6$ originated from the Hills
mechanism while the remaining $1/6$ accounts for other modes of ejection --
either a strong interaction of a single star with a binary, or the binary-binary
scattering. In some rare cases, the event list does not contain any information
about regularized interactions with other stars, which may in reality correspond
to any of the above mentioned mechanisms or even a close hyperbolic interaction
of two single stars in the tidal field of the SMBH. We excluded these stars
from further considerations. After the velocity correction to the Galactic potential,
we obtain on average $\approx12$ HVSs per realization.
\begin{figure}
\begin{center}
\includegraphics[width=\myfigwidth]{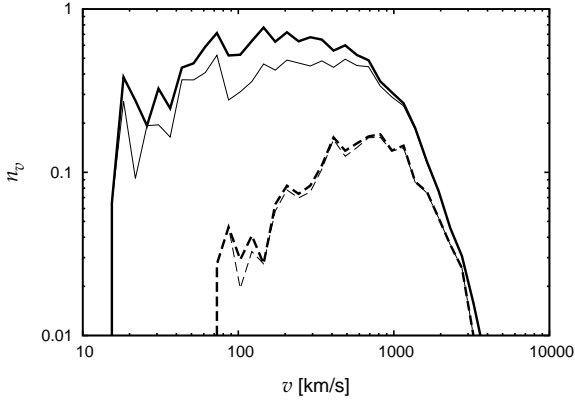}
\caption{\label{fig:hvs-vdist}
 Arbitrarily normalized velocity distributions of escaping stars detected at a
 distance of $8\,\pc$ from the SMBH within model {\modela} (solid lines). Dashed
 lines represent the velocity distributions of the HVSs that would be reached
 at a distance of $50\,\kpc$ from the Galactic center.
 Thick lines show velocity distributions of all escapers, while the thin ones
 correspond to the stars ejected via the Hills mechanism.}
\end{center}
\end{figure}
\begin{figure}
\begin{center}
\includegraphics[width=\myfigwidth]{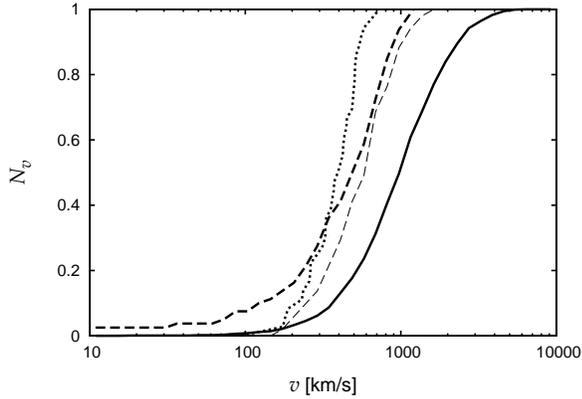}
\caption{\label{fig:hvs-vcdist}
 Cumulative distribution function of velocities of HVSs at $r=50\,\kpc$ from
 models {\modela} and {\modelb} are plotted with solid and dashed lines, respectively.
 Thin dashed line shows how the distribution function changes when deeper gravitational
 potential well of the Galaxy is assumed. Dotted line represents distribution
 function of Galactic rest-frame velocities of HVSs reported by \cite{Brown2014},
 including those declared as potentially bound to the Galaxy. The velocities are
 reduced to the same galactocentric distance of $50\,\kpc$ under the assumption
 that they move in the potential described by formula~(\ref{eq:galpot}).}
\end{center}
\end{figure}

\begin{figure}
\begin{center}
\includegraphics[width=\myfigwidth]{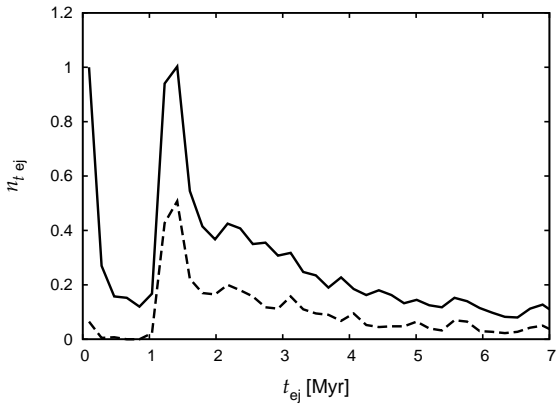}
\caption{\label{fig:hvs-tdist}
 Arbitrarily normalized distribution functions of the ejection times of all escaping
 stars in model {\modela} at $r=8\,\pc$ (solid lines) and the HVSs (dashed lines).}
\end{center}
\end{figure}
Figure \ref{fig:hvs-vdist} shows the distributions of the absolute values of the
velocities of all escaping stars at $8\,\pc$ and of their subset ejected via the
Hills mechanism within model {\modela} (thick and thin solid lines, respectively).
Comparison of the two
distributions clearly shows that the Hills mechanism dominates the ejection process for
$\vesc \gtrsim 500\,\kms$ at $r=8\,\pc$. Consequently, the potentially observable
HVSs at $r=50\,\kpc$ practically exclusively originate from the Hills mechanism.
Dashed lines in Figure~\ref{fig:hvs-vdist} represent expected velocity distributions
of the HVSs at $r=50\,\kpc$, i.e. distributions of velocities of the HVSs reduced
by motion in the Galactic potential~(\ref{eq:galpot}).

Model {\modela} appears to give velocity distribution considerably broader and
shifted to higher velocities with respect to the observed HVSs (see
Figure~\ref{fig:hvs-vcdist}). Figure~\ref{fig:hvs-vcdist}, however, also shows that
the velocity distribution of the HVSs strongly depends on the initial properties of the
stellar disc. Model {\modelb} which starts with all binaries initially relatively
weakly bound ($a_\mathrm{b} = 10\,\au$) gives considerably less HVSs with velocities
exceeding $1000\,\kms$ at $50\,\kpc$. This can be quite naturally
explained by the fact that the more tightly bound binaries from model {\modela}
get tidally disrupted closer to the SMBH, passing it with larger velocity which
is then inherited by the escaping star. Let us further note that distribution function of
binary semi-major axes in model {\modelb} evolved (broadened) quite rapidly due to
two-body relaxation. Typical semi-major axis of tidally broken-up binaries which
led to HVSs in model {\modelb} was about $1\,\au$.

Figure~\ref{fig:hvs-vcdist} also demonstrates that the distribution of velocities
of the HVSs is sensitive upon the shape of the Galactic potential. The two
distributions plotted for model {\modelb} (dashed lines) differ by the amount
of reduced kinetic energy -- the thick line corresponds to $\Delta E_\mathrm{K}
= 4\times 10^5 (\kms)^2$, according to formula~(\ref{eq:galpot}), while the thin
line corresponds to $\Delta E_\mathrm{K} = 5\times 10^5 (\kms)^2$. Note that
the simple model of Galactic potential~(\ref{eq:galpot}) may have similar level
of inaccuracy. For example,
while its parameters are such that it reasonably well fits mass of the Galactic
nuclear star cluster within the distance of $\approx 8\,\pc$ from the SMBH, it
very likely does not reproduce well the density distribution in that region and,
consequently, it underestimates depth of the Galactic potential. The way how
reduction of kinetic energy translates to change of velocities directly implies
that stars with lower velocities are affected more as it can be also inferred from
Figure~\ref{fig:hvs-vcdist}.

Considering large uncertainty in initial properties of the stellar disc together
with uncertainty of the depth of the Galactic potential, we are not able to
provide unique prediction on the velocity spectrum of the HVSs which may be
produced by the mechanism described in this paper. Some level of uncertainty
has to be also considered on the observational side when comparing models to
the observational data.

\begin{figure}
\begin{center}
\includegraphics[width=\myfigwidth]{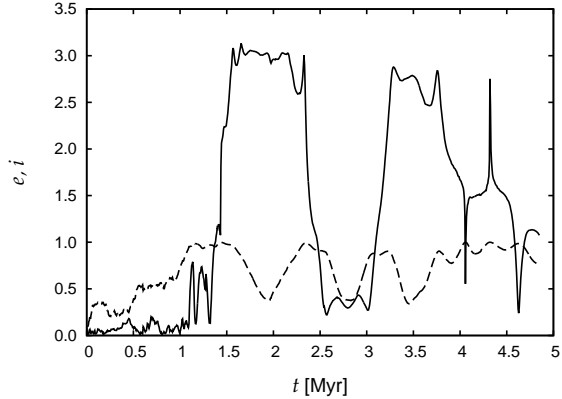}
\caption{\label{fig:star1815}
 Temporal evolution of inclination (solid line) and eccentricity (dashed) of an
 orbit from the inner part
 of the disc. Tidal break-up occurs at $t=1.15\,\myr$. From that time on, orbital
 elements of the component that remains bound to the SMBH are plotted.}
\end{center}
\end{figure}

In Figure~\ref{fig:hvs-tdist}, we plot the distribution function of the ejection
times of the escapers within model {\modela}.
Again, we distinguish distributions of all escapers and the HVSs.
We see an initial peak at $t=0$ which is especially apparent for the overall
distribution, i.e. it is dominated by the low velocity escapers.
These originate from those primordial
binaries that are immediately broken up in the tidal field of the SMBH. More
interesting is the relatively narrow peak of ejection of the HVSs which rises at
$t \approx 1\,\myr$ and slowly decays during the subsequent few million years. This
feature is a consequence of the coplanar eccentric Kozai-Lidov mechanism, which is
responsible for the production of the majority of the HVSs in our setup. 
Our results show that most of the oscillating orbits are located close to the inner
edge of the disc, thus having similar characteristic time-scale of the eccentric
Kozai-Lidov cycles. Hence, a large fraction
of the affected binaries reach high eccentricity at a common time. Typical
representative of the tidally disrupted binary is shown in Figure~\ref{fig:star1815}.
Initially, the lines represent orbital elements of the binary barycenter. At
$t = 1.15\,\myr$, eccentricity of its orbit around the SMBH reaches value $e > 0.99$
and the binary breaks up. From that time on, the lines in Figure~\ref{fig:star1815}
represent orbital elements of the component which remains bound to the SMBH, while
the other one escapes. Oscillations of orbital elements of the bound star shows
pattern typical for the coplanar eccentric Kozai-Lidov cycles (\citealt{Li14}; see also
\citealt{Haas15}) with
flips from corotation to counterrotation and vice versa. As the time proceeds,
pattern of the oscillations changes which is likely due to evolution of the source
of the perturbing potential (the stellar disc) in the {\nbody} environment.
Semi-major axis of the bound star just after the tidal break-up is $a \approx
0.015\,\pc$ which is approximately one half of that of the binary before that
event. The escaping star becomes unbound to the SMBH, having velocity of
$\approx 140\,\kms$ at a distance of $8\,\pc$ from the SMBH, i.e. this particular
star does not contribute to the population of the HVSs.

\begin{figure}
\begin{center}
\includegraphics[width=\myfigwidth]{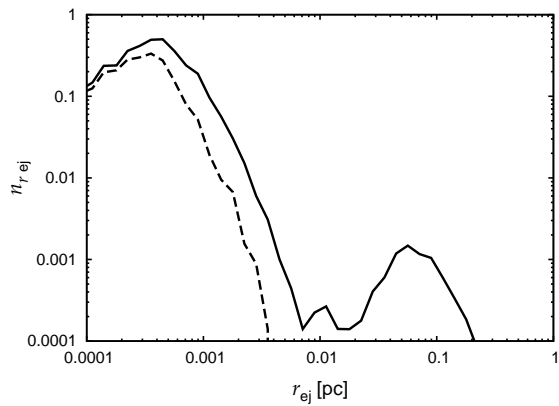}
\caption{\label{fig:hvs-rdist}
 Arbitrarily normalized distribution functions of the ejection radii of all escaping
 stars in model {\modela} at $r=8\,\pc$ (solid line) and the HVSs (dashed line).}
\end{center}
\end{figure}
Figure~\ref{fig:hvs-rdist} demonstrates that tidal break-ups of binaries within model
{\modela} occur in a relatively wide range of radii. This is caused by an initially wide
range of the (intrinsic) binary semi-major axes considered in our model which determine
how deep into the potential well of the SMBH the binary can penetrate before being tidally
broken up. We also see that HVSs practically
exclusively originate from radii $\lesssim 0.0005\,\pc$.

\begin{figure}
\begin{center}
\includegraphics[width=\myfigwidth]{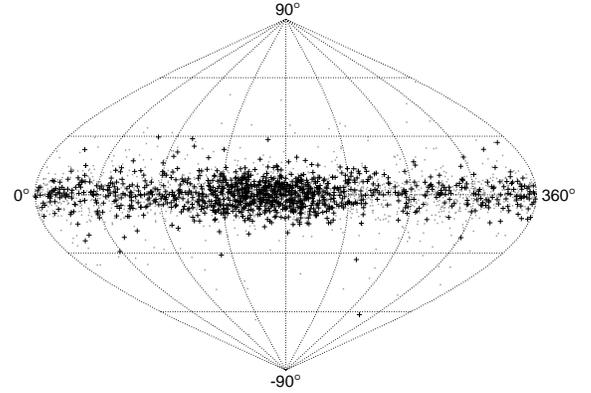}
\caption{\label{fig:hvs-directions}
 Distribution of the position vectors of the escaping stars projected in model {\modela}
 on a sphere
 centered on the SMBH; longitude is measured in the plane of the stellar disc; latitude
 is the angle measured from the stellar disc plane. Crosses represent HVSs, while the
 gray dots  stand for the rest of the escaping stars, i.e. those which would not reach
 galactocentric distances $\gtrsim 50\,\kpc$.}
\end{center}
\end{figure}
\begin{figure}
\begin{center}
\includegraphics[width=\myfigwidth]{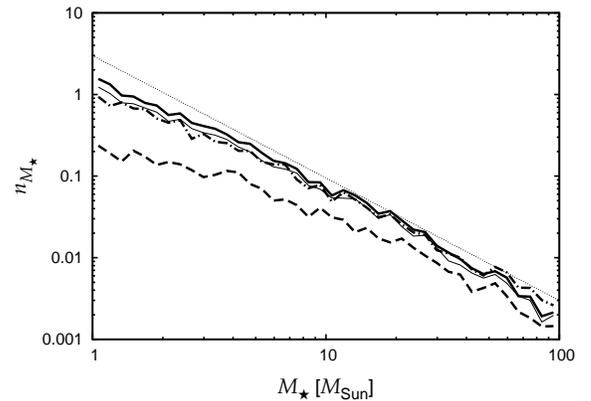}
\caption{\label{fig:mfunc}
 Mass functions of different groups of stars in model {\modela}: all escaping stars
 (thick solid line), all escaping stars ejected via the Hills mechanism (thin solid),
 HVSs (thick dashed) and the S-stars
 (thick dash-dotted). Thin dotted line represents the initial mass function of the disc.}
\end{center}
\end{figure}
\begin{figure*}
\begin{center}
\includegraphics[width=\textwidth]{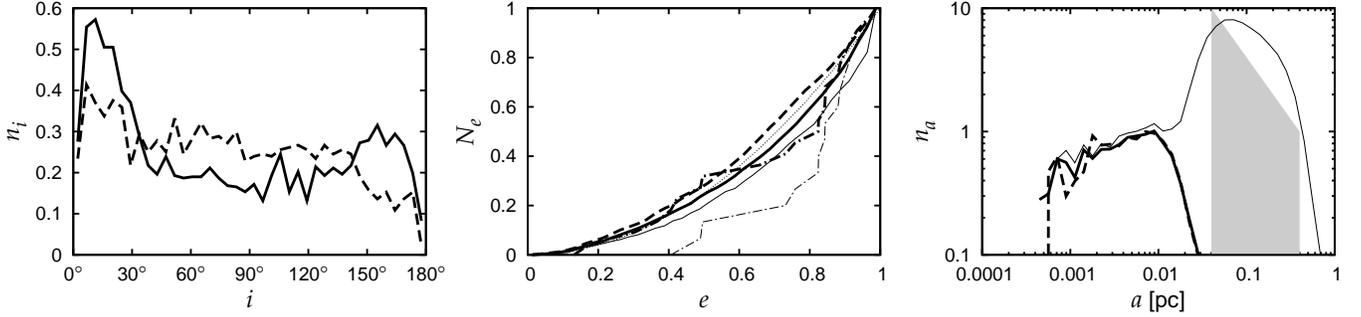}
\caption{\label{fig:ss-dist}
 Arbitrarily normalized distribution functions of inclinations and semi-major axes
 and a cumulative distribution function of eccentricities of the S-stars according
 to the model {\modela}. Thick solid lines show the distributions of the
 S-stars produced within the first $3.8\,\myr$ and they also represent the initial
 state of the follow-up model. Thick dashed lines describe the final state of the
 follow-up integration. Thin dotted line in the middle panel
 corresponds to thermal eccentricity distribution, $N_e \propto e^2$, thick and thin
 dash-dotted lines show eccentricity distribution of observed young stars in the
 Galactic center and of their subset with $a < 1\arcmin$, respectively. Thin solid
 line in the middle panel depicts eccentricity distribution within the model {\modelb}.
 Thin solid line in the right panel shows distribution function of semi-major axes of
 all stars within model {\modela} at $t = 3.8\,\myr$, \NEW{while the shaded area
 represents its initial state.}}
\end{center}
\end{figure*}
Figure~\ref{fig:hvs-directions} shows positions of the escaping
stars on the sphere in the sinusoidal projection. There is a clear
anisotropy in the distribution with majority of stars escaping along the plane of
the stellar disc. This is a natural consequence of the coplanar eccentric Kozai-Lidov
mechanism which is the main process that brings the binaries from the disc to extremely
eccentric orbits in our models.
In spite of the fact that the orbits change their inclination during the secular
evolution, this only happens hand-in-hand with evolution of eccentricities.
In particular, in the high-inclination phase ($i \approx 90\degr$), the eccentricity
is so large and argument of pericenter is close to $0$ or $\pi$ that
the orbit is embedded in the disc and the velocity vector of the star is nearly
parallel with the disc plane for major part of the orbit. It is then natural
to expect that the unbound component of a tidally broken up binary follows a trajectory
nearly parallel to the stellar disc. Beside the strong tendency of escaping
at low latitudes with respect to the disc, we also observe a significant
clustering of the ejected stars around a certain value of the
azimuthal angle in the plane of the disc. We attribute this to the
nature of the eccentric Kozai-Lidov mechanism which relies on specific mutual
orientation of the orbit and the eccentric perturbation of the central potential.
In spite of that the current observations are incomplete in sky-coverage, they
indicate statistically significant anisotropy of distribution of the HVSs on
the sky \citep{Brown2014}, which is in a qualitative accord with the outcome of
our model.

The currently observed HVSs are exclusively late B-type stars \citep[e.g.][]{Brown2014},
which is, nevertheless, just a selection effect and the mass spectrum of the
HVSs has to be considered unknown at this time. The numerical model, however,
enables us to predict the mass spectrum of the HVSs. Figure~\ref{fig:mfunc} shows
the mass function of the escaping stars produced during the whole integration of
model {\modela}. We see that it somewhat deviates from the initial mass function, in
particular, it is flatter at the low-mass end. This feature is more prominent for
the HVSs, i.e. it is primarily these stars that are
responsible for the deviation from the initial mass function. We see two possible reasons
for such a flattening of the mass function which are not mutually exclusive. First,
more massive binaries have on average a larger binding energy than the lighter
ones, i.e. they are able to sink deeper into the potential well of the SMBH
before tidally breaking up. Consequently, they achieve higher ejection
velocities, i.e. they have larger probability to be detected as HVSs. Second,
and a probably more important reason is that it is preferably the more massive
stars (binaries) which sink towards the SMBH within the parent disc due to two-body
relaxation. In spite of the fact that this process is not as pronounced as in
self-gravitating systems without a dominating central mass, our simulations show that
the mass function at the inner edge of the disc gets somewhat flatter already at
$t\approx 1\,\myr$. As the eccentric Kozai-Lidov cycles have a shorter period at
smaller radii in our settings, they preferentially push binaries from this region
to highly eccentric orbits on which they can tidally break up.
\subsection{S-stars}
\label{sec:sstars}
Let us now analyze the properties of the S-stars in our model as they were defined
in Section~\ref{sec:sstars-def}. On average we obtain $\approx 28$ S-stars per
realization of model {\modela} up to $t=7.6\,\myr$.\footnote{In $\approx 5\%$ of
the binary tidal break-ups, both components were accelerated above the escape
velocity from the SMBH.} In the subsequent paragraphs,
we will also qualitatively discuss comparison of the properties of the S-stars
in our model with those of the observed in the Galactic center. Note, however,
that such a comparison has to be handled with caution for two main reasons. First,
the number of the observed S-stars is quite small ($\approx 20$) which makes any
statistical analysis only marginally reliable. Second, we have strict definition of
S-stars in this paper based on their dynamical history. Such a definition cannot be
applied to the stars observed in the Galactic center. Actually, to our knowledge,
there is not any widely accepted definition of the observed S-stars. Therefore,
it is very difficult to compare the outcomes of numerical models with the observational
data.

Figure~\ref{fig:ss-dist} shows main statistical properties of the captured S-stars.
Left panel presents the distribution of their inclinations within model {\modela}.
We see two dominant peaks around $i=0$ and $\pi$, i.e.
corotating and counterrotating with respect to the parent disc. Similarly to
the distribution of the velocity vectors of the ejected stars, the high anisotropy
of the orientations of the orbits of the S-stars is a consequence of the
Kozai-Lidov mechanism.
Despite the vague definition of the observed S-stars, it is quite safe to
state that the young stars observed within the projected distance of
$\approx 1\arcsec \approx 0.04\,\pc$ from the SMBH have relatively
randomly oriented orbital planes (at least in comparison to the coherent orientations
of orbits of the stars forming the young stellar disc above this radius).
Hence, there has to be some process which randomizes the orbital orientations
in order to smear out their high anisotropy introduced by the process of their
formation discussed in this paper.
Indeed, it was suggested by various authors that the orientations of the
orbits of the S-stars may have undergone considerable evolution due to stellar dynamics.
In particular, resonant relaxation processes \citep{Rauch96} within the nuclear star
cluster were discussed e.g. by \cite{Hopman06}.
Another process capable of changing the orientations of the orbits is the
Kozai-Lidov oscillations due to a secondary, arbitrarily
inclined stellar disc \citep{Lockmann08}. A certain combination of the
two processes was considered by \cite{Chen14}. Our model
does not allow us to verify the resonant relaxation of the orbital parameters of the
S-stars as it does not include the spherical component of the nuclear star cluster
for numerical reasons. We tried to overcome this limitation by means of a follow-up
model. It consists of the S-stars whose initial kinematic state was taken from the
main model {\modela} at $t = 2\times 10^6 \unit{t} \approx 3.8\,\myr$, regardless of
the time of their formation. Motivation for this time is twofold. On one hand,
at $t\approx 3.8\,\myr$, on average a considerable number of S-stars
($\approx 21$) are already formed in the main model.
On the other hand we wish to test whether the
orbits of a substantial fraction of the S-stars can be randomized within a few
millions of years which is the estimated
life-time of the currently observed young stellar disc in the Galactic center.
The S-stars were embedded in a spherical cluster modeled by 500 stars of equal
mass ($1\,\msun$) with random orientations of their orbits, thermal distribution
of eccentricities and a distribution of the semi-major axes $\dist{a} \propto a^{1/2}$
in $\langle 0.002\,\pc,\;
0.02\,\pc \rangle$, i.e. within the domain of the S-stars. Distributions of the
orbital elements of the S-stars after $3.8\,\myr$ of the follow-up dynamical evolution
are plotted in Figure~\ref{fig:ss-dist} with the dashed lines. 
We observe a substantial evolution of the orientations of the orbits which tend
to become randomized.
Hence, we may state that our model of production of the S-stars via the Hills
mechanism is compatible with the orientations of the observed S-stars.

Distribution of eccentricities of the S-stars in our model {\modela} appears to be
close to the thermal one, $\dist{e} \propto e$ (Figure~\ref{fig:ss-dist},
middle panel) and it does not evolve considerably in time when embedded in the
spherical cluster. The distribution is roughly compatible with that of early-type
stars with determined orbital elements observed in the Galactic center (thick
dash-dotted line, data taken from \citealt{Gillessen09}).
The situation changes if only stars with semi-major axis smaller than $1\arcsec$
are considered (thin dash-dotted line in Figure~\ref{fig:ss-dist}).
\cite{Gillessen09} reported a super-thermal eccentricity distribution ($n_e \propto
e^{2.6\pm0.9}$) for this subset of stars in the Galactic center, which is
steeper than what we obtained within the considered models {\modela} and {\modelb}.
The apparent differences between the outcomes of our two models suggest, however,
that further variations of the initial setup may
lead to even more super-thermal distribution of eccentricities. Furthermore, in order
to obtain more realistic results, the numerical model would have to include all
resonant and relaxational processes for the whole integration time. Finally, the
comparison with the observational data needs to account for the problem of definition
of the S-stars mentioned above.

Showing the distribution of eccentricities of the captured stars, let us mention
that works of other authors \citep[e.g.][]{Perets09,Madigan11,Antonini2013}
discussing origin of the S-stars
often expect their eccentricities coming from the Hills mechanism to be exclusively
greater than $0.9$. Our results indicate that this assumption may not be valid
in the case when the orbit around the SMBH secularly evolves, i.e. the binary
becomes gradually more perturbed when passing closer and closer to the SMBH during
subsequent revolutions around it. Let us also remark that \cite{Hills1991} obtained extreme
values of eccentricities of captured stars only for highly radial orbits, while
larger impact parameters led to mean values of eccentricity considerably smaller
than $0.9$.

The S-stars produced by the Hills mechanism in model {\modela} occupy a rather wide range
of semi-major axes: $0.001\,\pc \lesssim a \lesssim 0.01\,\pc$ (see the right panel of
Figure~\ref{fig:ss-dist}). This is in accord with the observational data which
currently provide us with the smallest semi-major axis of $\approx 0.004\,\pc$
\citep[S0-102,][]{Meyer12}
while the outermost S-stars are located close to the inner edge
of the young stellar disc ($\approx 0.04\,\pc$). Comparison of the semi-major axis
distribution function of all stars and that of the S-stars in the right panel of
Figure~\ref{fig:ss-dist} indicates that the region below $0.01\,\pc$
is exclusively occupied by the S-stars in our model which justifies their formal
definition in Section~\ref{sec:sstars-def}. As expected, the
distribution of the semi-major axes did not change as their two-body relaxation time
is several orders of magnitude longer than the integration time.

Finally, we also evaluated the mass function of the captured S-stars in model {\modela}.
We found it to be similar to that of their former companions, i.e. the escaping
stars ejected via the Hills mechanism (see
Figure~\ref{fig:mfunc} and its description in the previous Section).
This result is not surprising for the initial setup of our model in which
the pairing of stars into binaries was biased towards equal masses.
\subsection{Disc structure evolution}
\label{sec:disc}
\NEW{On time-scale of the order of ten million years, both secular and relaxational
processes influence structure of the parent stellar disc. Due to relatively coherent
motions of the stars and, therefore, small initial velocity dispersion in the disc,
two-body relaxation is capable of altering its radial density profile, making it
flatter and broadening its extent both below and above its initial inner and outer edges,
respectively (see right
panel of Figure~\ref{fig:ss-dist}). Except for the population of S-stars,
which were transported below $0.01\,\pc$ due to the Hills mechanism, the radial
density profile does not differ significantly from that of models consisting of only
single stars presented in our previous works \citep{Subr14,Haas15}.

\begin{figure}
\begin{center}
\includegraphics[width=\myfigwidth]{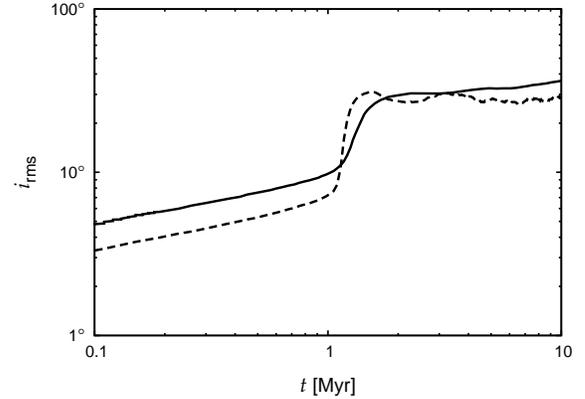}
\caption{\label{fig:irms}
 Temporal evolution of root mean square inclination of orbits from model {\modela}
 (solid line) and from model {\sf{}M8} of \citeauthor{Haas15} (\citeyear{Haas15};
 dashed).}
\end{center}
\end{figure}
Figure~\ref{fig:irms} shows that the two-body relaxation is also mainly responsible for
heating up the disc during the
initial phase of evolution ($\lesssim 1\,\myr$) which manifests itself by a power-law
growth of root-mean-square inclination of the orbits, $\irms \propto t^{1/4}$
\citep[e.g.][]{si2000}. At $t \approx 1\,\myr$, accelerated growth of $\irms$ is
observed, which is a consequence of relatively coherent Kozai-Lidov oscillations
of a subset of orbits which flip to counterrotation, thus pushing $\irms$ to
high values. The evolution of $\irms$ is qualitatively the same as that of model
{\sf{}M8} from \cite{Haas15} which has the initial global characteristics of the disc
(distribution of orbital elements and total mass) identical to the current model {\modela}.
The offset of the two models at $t \lesssim 1\,\myr$ we attribute to faster relaxation
of model {\modela} which, unlike the former model {\sf{}M8} has a broad mass spectrum.
Another additional source of heating of the disc in model {\modela} could be the
binaries which are not present in model {\sf{}M8}. Recent semi-analytical estimates
of \cite{Mikhaloff16} show, however, that binaries contribute only marginally to heating
of Keplerian stellar discs around SMBHs in comparison to the process of two-body
relaxation. Another kind of difference between evolution of $\irms$ of the two models
can be observed at $t\gtrsim1\,\myr$. Model {\sf{}M8} exhibits damped oscillations
of $\irms$ which is a consequence of longer
lasting coherence of the Kozai-Lidov oscillations of individual orbits. There are
at least two reasons why this feature is not observed in model {\modela}. First,
tidal disruptions of primordial binaries lead to abrupt changes of orbital elements which
affect the Kozai-Lidov cycles. Second, as argued above, model {\modela} undergoes
faster two-body relaxation due to the mass spectrum which gradually demotes the
eccentric perturbation. Both these arguments can also be used to explain higher value
of $\irms$ of model {\modela} at $t \gtrsim 3\,\myr$ which is likely due to
orbits that stay frozen in the counterrotating phase after interruption of their
Kozai-Lidov cycles.

\begin{figure}
\begin{center}
\includegraphics[width=\myfigwidth]{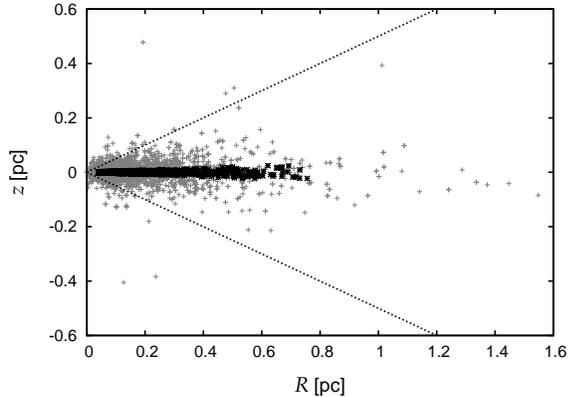}
\caption{\label{fig:disc_Rz}
 Azimuthal projection of position of stars from model {\modela} at $t=7.6\,\myr$
 (gray points) overplotted with initial state (black points). Dotted line indicates
 cone with half-opening angle $30\degr$.}
\end{center}
\end{figure}
In \cite{Haas15}, we already pointed out that $\irms$ is not straightforwardly
related to the disc geometrical thickness. Argumentation for this statement is
basically identical to that in
Section~\ref{sec:results_hvs} of the current paper: undergoing the coplanar Kozai-Lidov
oscillations, the orbits reach high inclination state with such values of eccentricity
and argument of pericentre that they are still embedded in a relatively thin disc
structure. Azimuthal projection of a snapshot of one realization of model {\modela}
at $t = 7.6\,\myr$
presented in Figure~\ref{fig:disc_Rz} confirms that except for several outliers, the
stellar disc is well confined within a cone with half-opening angle $\approx 30\degr$.}
\section{Discussion}
\label{sec:discussion}
We showed by means of a relatively simple model that both the S-stars and
the HVSs may have common origin in an eccentric stellar disc formed by fragmentation
of a gaseous cloud falling onto the SMBH.
Provided they were born as binaries, growth of the eccentricity of their orbits
around the SMBH due to the Kozai-Lidov mechanism induced by the disc itself
may have brought them to the tidal break-up radius at which they broke up,
leaving one component tightly bound to the SMBH and the other one
ejected away at a high velocity. Kinematic properties of the HVSs and the S-stars
in our model qualitatively agree with properties of the observed stars.
We have to keep in mind, however, that our
knowledge of these properties is incomplete due to the observational
limits and, therefore, a model matching such incomplete observational data may
actually be wrong. In this point, only improved observational data that are supposed
to be available in the future may help to test our model. Until then, there is
still room for improvements of the model to make it more realistic.

When, e.g., considering the stars as finite-size objects rather than point masses,
we may expect them to physically collide.
It was suggested by various authors that binary stars orbiting a SMBH may
merge due to the secular evolution of their (binary) eccentricity
\citep[e.g.][]{Antonini2010,Antonini2011,Prodan15,Stephan2016}. This event may
prevent the production of the HVSs provided the binary components merge before they
reach the tidal break-up radius. Whether or not the stellar collision occurs depends
on a complicated mixture of poorly constrained parameters
that determine the time-scales of the binary Kozai-Lidov oscillations and the
Kozai-Lidov cycles of the orbit around the SMBH. The fraction of binaries that merge
due to oscillations of the binary eccentricities varies from few per cent
\citep{Prodan15} up to $\sim10\%$ \citep{Stephan2016}.
It is, however, not clear, whether the binaries that produce HVSs in our model
would be affected on a similar level. In other words, the fraction of merging binaries
is likely to vary across the whole parameter space and we do not know, what is
its value in those parts which contribute to production of the HVSs.

Production rate of the HVSs and the S-stars could also change when another
sources of gravity capable of modifying the Kozai-Lidov dynamics are involved.
One of the observed and theoretically expected constituents of the Galactic nucleus
that was neglected in our main models
is the old component of the nuclear star cluster (which consists of late-type stars
and very likely also compact stellar remnants). Considering it to be spherically
symmetric to the first order of approximation, it generally has a tendency to damp
the Kozai-Lidov oscillations. This effect for the case of an axisymmetric perturbation
causing these oscillations was discussed, e.g., in \cite{Ivanov05}, \cite{Subr07}
and \cite{Karas07}; further extension for
the case of an eccentric perturbation, i.e. for the higher order Kozai-Lidov resonances,
can be found in \cite{Haas15}. In the latter paper, we demonstrated that with an
increasing mass of the spherical cluster, the higher order effects of the
Kozai-Lidov mechanism are damped first while the classical (quadrupole) cycles can
survive larger masses. The limiting mass of the cluster beyond
which the Kozai-Lidov mechanism is unable to push the stellar orbits to extreme
eccentricities depends on the mass distribution in both the cluster and the disc and also
on eccentricities of the stellar orbits. In particular, we found that a moderate
to high initial eccentricity of the disc is a key feature as it may lead to a global
angular momentum flow through the disc which is capable to push a subset of orbits
to higher eccentricities. Consequently, they can reach the resonant configuration
which leads to extreme Kozai-Lidov oscillations. Our numerical model in the current paper
is not suitable for a direct testing of the influence of the stellar cluster
on the production of the HVSs as addition of several tens of thousands of particles
would require much more computational time and, at the same time, it would lead to
a lower numerical stability. An alternate approach which is commonly used in similar studies, namely,
modeling the cluster by a smooth static potential, would require to model the
SMBH as an external potential as well, which, however, leads again to a lower numerical
stability of our integrations when the binary tidal break-ups occur.

\begin{figure}
\begin{center}
\includegraphics[width=\myfigwidth]{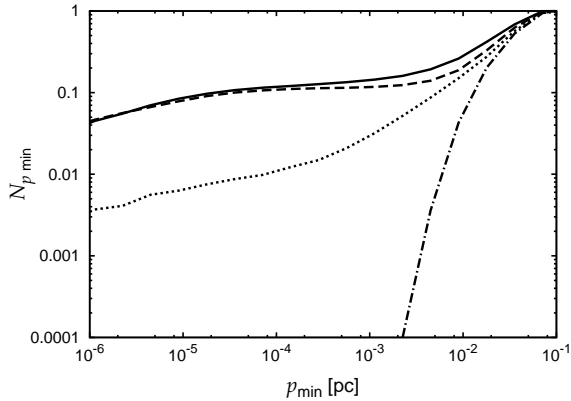}
\caption{\label{fig:perimin}
 Cumulative distributions of the minimal pericentral distance reached in the single-star models
 with the spherical cluster included. Solid, dashed, dotted and dash-dotted lines correspond
 to $\mc=0,\; 1\,\md,\; 4\,\md$ and $10\,\md$, respectively.}
\end{center}
\end{figure}

Here we make an attempt to estimate the rate of production of the HVSs and the S-stars
based on numerical integrations of models without the primordial binaries, which is
a numerically less complicated setup. The key idea lies in the empirical fact, that
majority of the tidal break-up events occurred below $0.0005\,\pc$ in our model {\modela}
(see Figure~\ref{fig:hvs-rdist}). Models without the primordial binaries
can then be used to determine the fraction of stars that reach the SMBH to a distance
smaller than the given value. Such a fraction can be
determined for systems embedded in the spherical potential as well as for the isolated
one. In particular, we integrated several models similar
to model {\modela} introduced in Section~\ref{sec:model_details} with the following
three key differences:
\begin{list}{}{\sublist{2em}}
\item[(i)] the SMBH is modeled as a fixed Keplerian potential
\item[(ii)] gravity of the spherical cluster is modeled by a fixed potential
$\Phi_\mathrm{c} \propto r^{3/2}$ which corresponds to a mass distribution
$\varrho_\mathrm{c}(r) \propto r^{-1/2}$. It has a single free parameter, $\mc$,
which determines the mass of the cluster enclosed within a radius $a_\mathrm{out}$
(the maximum initial value of the semi-major axis of the stellar orbits in the disc).
\item[(iii)] all stars are initially single.
\end{list}
We let these models with $\mc=0,\; 1\,\md,\; 4\,\md$ and $10\,\md$ to evolve for
$\approx 10\,\myr$ and monitored the minimal radial distance, $\pmin$, of the stars
from the SMBH reached during this time interval. Figure~\ref{fig:perimin} shows 
the cumulative distribution functions of $\pmin$ for the above described models
with different values of $\md$.
We see that the spherical cluster of a mass equal to the
mass of the disc leads to only small decrease of $\Dist{\pmin}$ at
$\pmin \approx 0.0005\,\pc$ with respect
to the isolated system. Therefore, we also do not suppose the number of HVSs to be
altered considerably in this case. For $\mc = 4\,\md$, the number of stars plunging
below $0.0005\,\pc$ is lowered by a factor of $\approx 5$ with respect to the case
without the spherical cluster included and we expect the number of the HVSs to be
lowered by a similar factor, i.e. to $\approx 2$ in contrary to $\approx 12$ obtained
for the isolated model (see Section~\ref{sec:results_hvs}). Finally, the spherical
cluster of mass $\mc = 10\,\md$ practically completely damps the Kozai-Lidov
oscillations and we do not expect any HVSs or S-stars to be formed via the Hills mechanism
within the models described in this paper in such a case (see, however, \citealt{Haas15}
for discussion of other initial setups in which considerable amount of oscillating orbits
occurred even for $\mc \approx 10\,\md$). Note that the distribution of $\pmin$ does
not provide us directly with the expected number of the HVSs as not all particles
reaching the critical value would still be in the form of binaries even in the case of
a 100\% initial binary fraction and the argumentation above relies on comparison
of the isolated models with and without the primordial binaries.

Not only total number of the HVSs and the S-stars formed by the Hills mechanism
induced by the Kozai-Lidov oscillations of binary stars orbits around the SMBH could
be affected by presence of the spherical stellar cluster. The strong anisotropy
of the ejected stars as well as the temporal burst of their production described
in Section~\ref{sec:results_hvs} are features of the coplanar eccentric Kozai-Lidov
mechanism. Increasing mass of the stellar cluster first damps this mode \citep{Haas15}
and, therefore, we suppose its imprints on the HVSs velocity directions and times
of ejection to be suppressed as well. The classical (quadrupole) Kozai-Lidov
oscillations are also likely to bring stellar orbits to highly eccentric states with
specific orientations of the orbital planes, in this case, nearly parallel to the
plane of the disc.
Nevertheless, there is no reason to expect preferred direction of the eccentric
vector in this plane. In \cite{Haas15} we showed that these classical oscillations
occur from the very beginning of the disc evolution till the end of our integrations,
thus they are supposed to suppress the dominant peak in the distribution of ejection
times of the HVSs. Single-star
models introduced above are not suitable for determination of properties
of escaping stars. Therefore, complex models including large fraction of primordial
binaries in the disc as well as gravity of the spherical cluster deserve to be a subject
of future investigations.

There is growing evidence for tidal disruptions of stars in the vicinity of
SMBHs in distant galaxies \citep[see e.g.][]{Saxton12,Komossa15}. An interesting
side-effect of the tidal break-ups of binaries on orbits undergoing the Kozai-Lidov
oscillations is that it may prevent the individual stars from being tidally disrupted.
The Kozai-Lidov oscillations change the orbital elements
of orbits around the SMBH regardless of the multiplicity of the orbiting system. At the
maximum of eccentricity, the pericenter may lie below the stellar
tidal radius, i.e. if the system were formed by a single star, it would be
disrupted. However, a binary tidally breaks up before
it reaches the stellar tidal radius. Due to that, one star is
ejected away, thus being saved from tidal disruption, while the other one
remains parked on a tightly bound orbit whose pericenter is, however, typically
well above the stellar tidal radius. Orbit of the bound star may not further
undergo Kozai-Lidov oscillations, as it has abruptly changed its orbital elements
during the binary tidal disruption event. Hence, it may also be saved from being
tidally disrupted.

Finally, let us recall that the process discussed in this work led to a substantial
number of stars ejected from the stellar disc with velocities only marginally
exceeding the escape velocity from the SMBH. These stars should be found in the
Galactic bulge. Interestingly, several young stars at distances of the order of
tens of parsecs from the {\sgra} are observed. Their origin is unclear.
Some of them may have been ejected from the young star clusters Arches and Quintuplet.
However, it has been discussed by \citet{Habibi14} that these two clusters could
not be birth places of all of them. Let us close our discussion with a suggestion
that also the young stellar disc may have contributed to this population of
young stars.
\section{Conclusions}
\label{conclusions}
We have investigated by means of direct {\nbody} integrations the production of
the HVSs and the S-stars from an eccentric stellar disc around the SMBH through
tidal break-ups of binaries brought to the vicinity of the SMBH via the coplanar
eccentric Kozai-Lidov mechanism induced by the stellar disc itself. In accord with
the principle of the Occam's razor, this model of the origin of the HVSs
and the S-stars observed in the Galactic halo and nucleus, respectively, relies
on as small number of constituents as possible. At the same time, however, it
assumes recurrent star formation episodes to occur in
the Galactic center. The currently observed HVSs required at least $50\,\myr$
to travel from the Galactic center to the halo, i.e. they can not
originate from the young stellar disc observed now
in the Galactic center which is $\lesssim 7\,\myr$ old \citep[e.g.][]{paumard2006,Lu13}.

Velocity distribution of the HVSs in our model as well as their total number
strongly depends on the properties of the initial binary population. Unfortunately,
this piece of information is unavailable for the case of the young stellar disc
observed in the Galactic center, which otherwise, is considered to be a template
configuration. Therefore, any strong statement on the viability of our model
based  on the observational data cannot be made at this moment. The most
prominent `smoking gun' of our hypothesis appears to be strong anisotropy of
distribution of the HVSs on the sky which is a robust feature of the model and is
in accord with the current observational data. Another
characteristic feature of the HVSs in our model is a burst-like mode of their
formation. This feature is neither confirmed, nor excluded for the currently observed
HVSs \citep{Brown2014}. Future observations may bring stronger constraints in this
point. Note however, that (i) in spite that our model predicts one dominant peak
of ejection, still large fraction of HVSs are produced in wide time range and (ii)
more sophisticated model including spherical cluster is likely to lower the dominant
peak.

Hand in hand with production of the HVSs, our model of tidal break-ups of binary
stars originating from an eccentric disc also leads to transportation of
stars to orbits tightly bound to the SMBH within a few millions of years.
These could be then observed as S-stars. Strong initial anisotropy of normal vectors
of their orbital planes is very quickly (within another few millions of years)
smeared out due to resonant relaxation within an embedding stellar cluster. Hence,
unlike in the case of the HVSs, we do not observe any characteristic feature
of the distribution functions of the orbital elements of the S-stars formed in
our model which could serve as a strong test bed of its relevance. Finally, let us
note that quite rapid formation of the population of the S-stars from an eccentric
disc implies that, according to our model, some of the currently observed
S-stars may have originated from the young stellar disc observed in the Galactic
center nowadays. On the other hand, life-time of the observed S-stars allows them
to be original companions of the HVSs currently observed in the Galactic halo,
i.e. originating from some previous star formation episode. The two scenarios
are not mutually
exclusive. Actually, we estimate the number of S-stars originating from the currently
observed young stellar disc in the Galactic center to be only a few when considering
damping role of the spherical component of the nuclear star cluster.

\NEW{Beside the properties of the HVSs and the S-stars, we also evaluated evolution
of the parent stellar disc. We found that it evolves in a way similar to models without
primordial binaries. In particular, the binaries do not contribute significantly
to heating of the disc, i.e. the model of isolated disc is not able to reproduce
spatial distribution of all the young stars in the Galactic center, substantial
fraction of which are observed well above the plane of the
young stellar disc (CWS). The hypothesis
of common origin of all young stars observed in the GC from a single parent
disc may be kept viable if we assume some other perturbing source(s) of gravity
to be present (see, e.g., \citealt{ssk2009,hsk2011,hsv2011} for discussion of
influence of an outer massive gaseous torus on evolution of the stellar disc).}
\section*{Acknowledgments}
We thank the anonymous referee for helpful comments.
L.\v{S}. acknowledges support from the Grant Agency of the Czech Republic via Grant
No.~GACR 14-10625S and from the National Science Foundation under Grant
No.~PHYS-1066293 and the hospitality of the Aspen Center for Physics. J.H. was
supported by Charles University project UNCE-204020.

\end{document}